\documentclass[useAMS,usenatbib]{mn2e}

\usepackage{amsmath}
\usepackage{epsfig,graphics}
\usepackage{rotating} 
\usepackage{epsfig}

\usepackage{graphicx,natbib,multirow,topcapt}
\usepackage{savefnmark}
\usepackage[perpage,symbol*]{footmisc}
    
\newcommand {\be}{\begin{equation}}
\newcommand {\ee}{\end{equation}}  
\renewcommand{\theenumi}{\arabic{enumi}.}

\newcommand{\lesssim}{\hbox{\rlap{\hbox{\lower4pt\hbox{$\sim$}}}\hbox{$<$}}}
\newcommand{\gtrsim}{\mathrel{\hbox{\rlap{\hbox{\lower4pt\hbox{$\sim$}}}\hbox{$>$}}}}  

\newcommand{\NH}{N_{\rm{H}}}

\title[Detection of GRB afterglows with \textit{SRG}/eROSITA]
{\textit{SRG}/eROSITA prospects for detection of GRB afterglows} 

\author[I. Khabibullin et al.]{I. Khabibullin$^{1}$\thanks{E-mail:
khabibullin@iki.rssi.ru}, S. Sazonov$^{1}$ and R. Sunyaev$^{2,1}$\\
$^{1}$Space Research Institute, Russian Academy of Sciences,
Profsoyuznaya 84/32, 117997 Moscow, Russia\\
$^{2}$Max-Planck-Institut f\"ur Astrophysik,
Karl-Schwarzschild-Str. 1, 85740 Garching bei M\"unchen, Germany
}

\begin{document}

\maketitle

\begin{abstract}

We discuss the potential of the eROSITA telescope on board the
\emph{Spectrum-X-Gamma} observatory to detect gamma-ray burst (GRB)
X-ray afterglows during its 4-year all-sky survey. The expected
rate of afterglows associated with long-duration GRBs without any
information on the bursts proper that can be identified by a
characteristic power-law light curve in the eROSITA data is
4--8~events per year. An additional small number, $\lesssim 2$~per
year, of afterglows may be associated with short GRBs, ultra hard (GeV)
GRBs and X-ray flashes. eROSITA can thus provide the first unbiased
(unaffected by GRB triggering) sample of $\lesssim 40$ X-ray
afterglows, which can be used for statistical studies of GRB
afterglows and for constraining the shape of the GRB $\log N$--$\log
S$ distribution at its low-fluence end. The total 
number of afterglows detected by eROSITA may be yet higher due to orphan
afterglows and failed GRBs. The actual detection rate could thus
provide interesting constraints on the properties of relativistic jets
associated with collapse of massive stars. Finally, eROSITA can
provide accurate ($\lesssim 30''$) coordinates of newly discovered
afterglows within a day after the event, early enough for scheduling
further follow-up observations.

\end{abstract}

\section{Introduction}

The main objective of the \emph{Spectrum-Roentgen-Gamma} (\emph{SRG}) 
observatory is to perform a sensitive all-sky survey in the
0.3--12 keV energy band with the eROSITA\footnote{Extended ROentgen Survey with an Imaging Telescope Array} \citep{Predehl11} and ART-XC\footnote{Astronomical Roentgen Telescope -- X-ray Concentrator}
\citep{Pavlinsky11} telescopes. The survey (see \S\ref{s:tasks} for
details) will last 4 years and consist of 8 repeated complete scans of
the sky. The telescopes will be scanning the sky in great circles 
as a result of the spacecraft's rotation with a period of 4~hours 
around its axis pointed at the Sun. This
observational strategy provides the possibility of studying variable
and transient X-ray sources on three characteristic time-scales
corresponding to 1) the duration of a single scan of a point source,
$\lesssim 40$~s, 2) the duration of a single observation of a source,
$\gtrsim 1$~day (consisting of $\gtrsim 6$ consecutive rotation
cycles, depending on the ecliptic latitude), and 3) the duration of
a single all-sky scan (6~months).

Information on temporal behaviour can greatly assist in identifying
the types of X-ray sources discovered during the eROSITA all-sky
survey. In particular, the X-ray afterglows of cosmic gamma-ray bursts  
(GRBs) form a class of bright transient sources that usually
demonstrate power-law decay during the first hours and days after the  
burst (see \citealt{gehrels09} for a review). Therefore, GRB
afterglows can manifest themselves by a distinct variability pattern
on the time-scale of several successive eROSITA scans, which in
principle makes it possible to identify such events by analysing the
X-ray light curves of sources detected during the all-sky survey. 

The purpose of this paper is to estimate the detection rates of GRB
X-ray afterglows during the eROSITA all-sky survey. Previously,
\cite{Greiner00} carried out a search for afterglows of untriggered
GRBs in the \textit{ROSAT} all-sky survey (RASS) data and found 23
afterglow candidates. However, a closer examination indicated that at
least half and perhaps the majority of these events were flares of
late-type stars. Taking into account eROSITA's better (by a factor of
$\sim 4$ in the 0.5--2~keV energy band) sensitivity and a factor of 2
larger sky coverage (survey duration times field of view area) of the
planned survey, we can expect a significantly larger number of
detected GRB afterglows. We note in passing that in addition to
afterglows, the \emph{SRG} all-sky survey may also detect a significant
number of GRBs themselves, depending on the unknown shape of the
$\log N$--$\log S$ distribution of GRBs at very low fluences. This topic
will be addressed elsewhere.

The main motivation for such a study is that eROSITA can provide the
\emph{first unbiased sample of X-ray afterglows}. The problem
with all existing samples of afterglows is that they are based on
trigerred GRBs and thus determined by the energy range, sensitivity
and strategy of the particular GRB experiments. As we discuss in this
paper, eROSITA may find a significant number of afterglows
associated with GRBs falling near or below the detection threshold
and/or having the spectral maximum outside the energy range of existing
GRB monitors. This will make it possible to construct an unbiased
distribution of X-ray afterglow fluxes and will also provide valuable
constraints on the shape of the GRB $\log N$--$\log S$ distribution at
low prompt emission fluences. In addition, eROSITA may find a
significant number of 'orphan' afterglows, i.e. events without
preceding prompt high-energy emission
(e.g. \citealt{Rossi2002,Nakar2003}), and afterglows associated with
'failed GRBs' (e.g. \citealt{Huang2002}), which will provide
interesting constraints on the properties of relativistic jets
produced during the collapse of massive stars \citep{MW1999}. 

\section{Tasks and analytical estimates} 
\label{s:tasks}

The \emph{SRG} observatory will be launched into the L2 point of the
Earth--Sun system and the first 4 years of the mission will be
devoted to performing an all-sky X-ray survey. The satellite will be
rotating with a period of $T=4$~hours around its axis pointed at the
Sun (or a few degrees away from the Sun, the exact strategy is still
to be decided), with the telescopes observing the sky at right angles 
to the axis. Consequently, the rotation axis will be moving at a
speed of 1~deg per day following the orbit of the L2 point around the
Sun. As a result, a complete survey of the sky will be done every
6~months and a total of 8~scans will be completed over the course of
the mission. With the 1~deg-diameter field of view (FoV), eROSITA will
be scanning the sky with a speed $\frac{dS}{dt}\approx
\frac{S_0}{180T}\approx 1$~deg$^2$ per minute (where $S_0$ is the
all-sky area). A typical position on the sky will pass through the FoV
during 6 consecutive spacecraft rotations once during each all-sky
scan. A single scan of a point source will last $\simeq 40$~s, the
time it takes for the source to cross the FoV through its
centre. Therefore, most of the sky will receive an exposure $\sim
2$~ks by the end of the survey. In reality, the received exposure
gradually increases towards the ecliptic poles, so that the ecliptic
caps will receive up to $\sim 30$~ks by the end of the survey
\citep{Pavlinsky12}. This also means that any sources located in
these regions will transit the eROSITA FoV during more than 6
successive rotation cycles.

Below we consider two scientifically interesting tasks for the
\emph{SRG}/eROSITA survey: i) identification of afterglows without any
information about the bursts proper and ii) search for afterglows of
triggered (by instruments aboard \emph{SRG} or other observatories)
GRBs. 

\subsection{Task 1: identification of afterglows of non-triggered GRBs}
\label{s:task1}

\emph{Swift}/XRT observations \citep{Nousek06,Zhang06} have
demonstrated that the light curves of GRB X-ray afterglows cannot
always be described by a power law and frequently not even by a
combination of several power laws (usually this is the case when one
or more flares are observed), as had been previously suggested by
\emph{BeppoSAX} observations of late-time afterglows
\citep{dePasquale06}. However, eROSITA should be most efficient for
studying GRB afterglows during the period from $\sim 10^3$ to $\sim
10^5$~s after the prompt emission, which usually corresponds to the
phase of smooth decay with a slope of approximately 1.2 (see segment
III in the cartoon in \citealt{Zhang06}). We thus base our treatment
below on the requirement that GRB X-ray afterglow candidates have
power-law-like light curves.  

\begin{figure}
\centering
\includegraphics[width=\columnwidth]{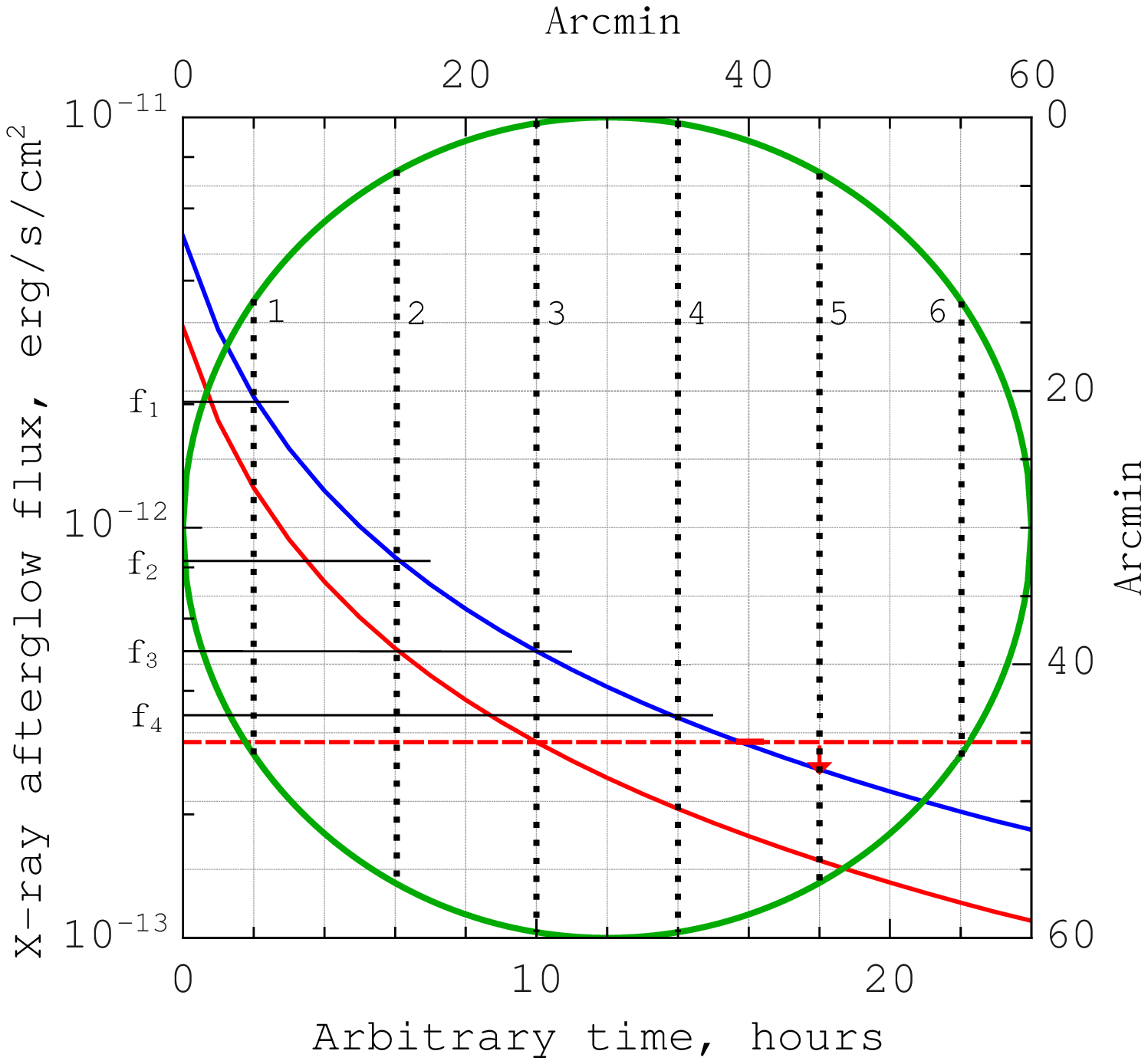}
\topcaption{Illustration of a typical GRB X-ray afterglow observation
  by eROSITA. Here, the GRB source passes through the eROSITA FoV
  (1~deg-diameter green circle with the corresponding right and top
  axes) during 6 consecutive scans, these transits being shown by
  vertical black lines. As a result, the afterglow light curve (blue
  line) is measured at 6 instants separated by 4~hours (see the
  corresponding left and bottom axes), but only the first 4
  measurements exceed the detection limit (horizontal red
  line).}
\label{obs}
\end{figure}

Consider the light curve of a GRB afterglow in the 0.5--2 keV energy
band (where eROSITA is most sensitive) of the form  
\begin{equation}
	F_{X}(t)=F_{X,12}\left(\frac{t-t_{0}}{12~{\rm
            hr}}\right)^{-\delta}=F_{X,12}\left(\frac{\tau}{12~{\rm
            hr}}\right)^{-\delta},
\label{decay}	
\end{equation}
where $F_{X,12}  $ is the X-ray flux at 12~hours after the prompt emission
and $\tau=t-t_{0} $ is the 'age' of the afterglow, i.e. the time since
its onset (in hours). Such an afterglow can be detected by eROSITA
until the time 
\begin{equation}
\tau_{e}(F_{X,12},\delta)=12~{\rm hr}
\left(\frac{F_{X,12}}{F_{e}}\right)^{1/\delta}, 
\end{equation}
where $F_{e}  $ is the eROSITA detection limit for one ($\simeq 40$~s 
long) scan of a point source. If $ \tau_{e}\geq n T $, a light curve
containing up to n+1 points can be obtained under suitable spacecraft
orientation conditions. In principle, three consecutive flux
measurements, $f_{1}$, $f_{2}$ and $f_{3}$, are sufficient to
reconstruct a power-law-shaped light curve (see
Fig.~\ref{obs}). Indeed, in this case we have three equations for
determining three parameters: $F_{X,12}$, $\delta$ and $t_{0}$, or 
equivalently, $F_{X,12}$, $\delta$ and $\tau_{1}$, where $\tau_{1}$ is
the age of the afterglow during the first successful
scan. Specifically, the decay index $\delta$ can be found by solving
the equation
\begin{equation}
2\left(\frac{f_{1}}{f_{2}}\right)^{1/\delta}-\left(\frac{f_{1}}{f_{3}}\right)^{1/\delta}=1,   
\label{delta}
\end{equation}
making further determination of $\tau_{1}$ and $F_{X,12}$
trivial. However, in reality $f_{1}$, $f_{2}$ and $f_{3}$ are the 
\emph{measured} values of the afterglow flux, and it is also necessary
to take into account the uncertainties associated with flux
measurement. 

We thus propose as the criteria for identification of afterglow
candidates that: 
\begin{enumerate} 

\item
There is a sequence of 3 successive scans with the measured fluxes
($f_1$, $f_2$ and $f_3$) exceeding $F_e$;

\item
The $f_1$, $f_2$ and $f_3$ fluxes are consistent within their
uncertainties with a power-law decay with $0.5<\delta<3.5$ (see Appendix 1);

\item
The fluxes measured in scans preceding the $f_1$ measurement, if there
are any, are consistent with zero (to ensure that there is no rising
phase that would contradict a GRB afterglow origin); 

\item
The fluxes (or upper limits) measured in scans succeeding the $f_3$
measurement, if there are any, confirm the power-law decay.

\end{enumerate}

The second of the conditions listed above, together with the
characteristics of eROSITA telescope, implies (see Appendix~1) 
that the effective limit ($F_e$) for detection and identification of
GRB afterglows lies between $2\times 10^{-13}$ and $3\times
10^{-13}$~erg~s$^{-1}$~cm$^{-2}$ (0.5--2~keV). Here, the higher value
is more conservative, as it allows for the possibility of a
significant background contribution to the detected photon flux  
and some uncertainty in the characteristics of the eROSITA telescope;
we will use both values for our estimates.

Thus, for confident identification of a candidate afterglow we need at
least three consecutive detections above the threshold specified
above. This suggests that only afterglows
with $\tau_{e}\geq 2T=8$~hr are useful for our purposes. Furthermore,
such afterglows can only be suitable if detected for the first time
within $\tau_{m}=\tau_{e}-2T \geq 0$ after the onset of the
afterglow. In other words, $\tau_{m}$ defines the 'time depth' of an
interesting afterglow. For a given afterglow, we can further introduce
a survey volume $d\epsilon=\tau_{m}dS$, where $dS$ is the survey area,
and a survey rate $\frac{d\epsilon}{dt}=\tau_{m}\frac{dS}{dt}$. As was
mentioned above, the eROSITA scanning speed
$\frac{dS}{dt}=\frac{S_{0}}{180T}$. However, because eROSITA will scan
a given position in the sky typically for a total of 6 consecutive
rotation cycles, only half of the FoV (see Fig.~\ref{obs}) is actually
suitable for producing the first point of an at least three-point
light curve. Furthermore, only 1/3 of this area has the full time
depth $\tau_{m}$\footnote{Strictly speaking, the time depth of this
  area is $\min(T_{0},\tau_{m})$, where $T_{0}\approx 180\times 24$~hr
  -- the eROSITA all-sky scan period, but because the rate of GRBs
  with $\tau_{m}>T_{0}$ is extremely small, the associated error is
  negligible.}. For the remaining two thirds, the time depth equals
$\min(T,\tau_{m})$, because a given event could already be detected in
the FoV $T=4$~hr before. Thus, effectively 
\begin{equation}
 \frac{d\epsilon}{dt}\left(\tau_{m}\right)=\tau_{m}\frac{1}{3}\cdot
 \frac{1}{2}\frac{S_{0}}{180 T}+\min(T,\tau_{m})\frac{2}{3}\cdot
 \frac{1}{2}\frac{S_{0}}{180 T}.  
 \label{svol}
\end{equation}
Let $r\left(\tau_{m}\right)$ be the probability density function of
$\tau_{m}$ for GRB afterglows, normalised to some total GRB rate per
unit solid angle, $R=\int r\left(\tau_{m}\right)d\tau_{m}$. Then, the
expected number of GRB afterglows detected (and identified as such) by 
eROSITA during a unit time interval is  
\begin{equation}
N_1= R\langle\frac{d\epsilon}{dt}\rangle,
\end{equation}
where 
\begin{equation}
\langle\frac{d\epsilon}{dt}\rangle=\frac{1}{R}\int
r\left(\tau_{m}\right)\frac{d\epsilon
}{dt}\left(\tau_{m}\right)d\tau_{m}. 
\end{equation}
Therefore, to estimate the expected number of GRB afterglows, one
needs to specify the function $r\left(\tau_{m}\right)$, which can 
in principle be derived from observed distributions of GRB 
fluences, afterglow X-ray fluxes and decay indices. 

Before proceeding to accurate estimates, it is useful to make some
simplifying assumptions and determine a lower limit for the expected 
number of GRB afterglows analytically. Due to the form of
equation~(\ref{svol}), it is natural to divide all potentially
interesting afterglows into three categories:
\begin{enumerate}

\item
$ \tau_{e}< 8$~hr, hence $ \tau_{m}= 0 $ and $\frac{d\epsilon}{dt}=0$,

\item
$8~{\rm hr}\leq\tau_{e}< 12~{\rm hr}$, hence $ 0\leq\tau_{m}<
4$~hr and $\frac{d\epsilon}{dt}=\tau_{m}\frac{S_{0}}{360T}$,

\item
$\tau_{e}\geq 12$~hr, hence $\tau_{m}\geq 4$~hr and
$\frac{d\epsilon}{dt}=\frac{\tau_{m}+2T}{3}\frac{S_{0}}{360T}=\frac{\tau_{e}}{3} 
  \frac{S_{0}}{360T}$.  
\end{enumerate}
The \emph{Swift}/XRT sample of GRB afterglows is consistent with
a lognormal distribution of $F_{X,12}$ (2--10~keV) with
$<F_{X,12}> \approx 3\times 10^{-13}$~erg~s$^{-1}$~cm$^{-2}$ and 
$\sigma=0.5$ \citep{berger05}. Assuming 
a Crab-like X-ray spectrum with a photon index $\Gamma=2$ and fixing
the light curve slope at $\delta=1.3$, we find that for $F_{e}=3\times
10^{-13}$~erg~s$^{-1}$~cm$^{-2}$ (0.5--2~keV), approximately half of
all \emph{Swift}/XRT GRB afterglows have $\tau_{e} \geq 12$~hr and
thus fall into the third category described above. Taking into
account only such (bright) events, the total rate of afterglows
detected by eROSITA will be
$N_1>\frac{R}{2}\frac{<\tau_{e}>}{3\cdot360T}$, where
$<\tau_{e}>=12~{\rm hr}\frac{<F_{X,12}^{1/\delta}>}{F_{e}^{1/\delta}}$  
is the average over the lognormal distribution of fluxes for
$F_{X,12}$ (0.5--2~keV) $>F_{e}$. Given that $T=4$~hr and assuming a
total rate of GRBs $R\approx 1000/S_{0}$ per year (which corresponds to the
\textit{Swift} detection rate recalculated to the full-sky
area \citealt{dai09}), we obtain that  
\begin{equation}
N_1 >3.4~{\rm per}~{\rm year}.
\end{equation}
This is a conservative lower limit on the total afterglow detection
rate by eROSITA. In reality, the detection rate should be higher owing
to the contribution of weaker afterglows ($F_{X,12}<F_{e}$), but the
result should be sensitive to the shape of the $\log N$--$\log S$
distribution of GRB fluences. We obtain more accurate estimates in
\S\ref{s:simul} using Monte Carlo simulations. 

\subsection{Task 2: afterglows of triggered GRBs}
\label{s:task2}

A different situation arises if a GRB monitor provides sufficiently
accurate coordinates of a GRB. In this case, just one detection 
will be sufficient to identify its X-ray afterglow and measure its
flux at a known time after the prompt emission. The effective eROSITA 
detection limit for this task is $F_d= 1\times
10^{-13}$~erg~s$^{-1}$~cm$^{-2}$ (see Appendix~2), i.e. somewhat lower
than for Task 1 discussed above. 

In the considered case, $ \tau_m=\tau_e $ and the whole FoV is
suitable for the first (and the only needed) detection of an
afterglow: 1/6 of the FoV has time depth $\tau_e$ and the other 5/6 of
the FoV $\min (T, \tau_e)$. Consequently, 
\begin{equation}
 \frac{d\epsilon}{dt}\left(\tau_{m}\right)=\tau_{e}\frac{1}{6}\cdot
 \frac{S_{0}}{180T}+\min(T,\tau_{e})\frac{5}{6}\cdot
 \frac{S_{0}}{180T}.  
 \label{svol2}
\end{equation}
Similar to Task 1, we can find a lower limit on the eROSITA afterglow
detection rate for Task~2 using information on \emph{Swift} X-ray afterglows:
\begin{equation}
N_2 > 15~{\rm per}~{\rm year}.
\end{equation} 
This estimate is based on the optimistic assumption that at the time of the \emph{SRG} mission there is a GRB monitor or collection of GRB monitors that constantly cover the whole sky with sufficient senstivity (corresponding to $ \sim 10^{-8} erg  ~ s^{-1} ~ cm^{-2}$ in 15-150 keV energy band ) and localization accuracy ($ \sim $ 10 arcmin). For example, \emph{Swift}/BAT provides such characteristics except that its sky coverage is only 1.4 sr (half-coded).


Finally, we note the interesting possibility of searching with eROSITA for afterglows of GRBs detected by the Interplanetary Network (IPN, \citealt{Hurley2011}). IPN events are often localised (by triangulation) to a large annulus on the sky. The presence of a decaying X-ray source at the intersection of the eROSITA FOV with such an annulus within the first hours--days after a GRB would indicate the detection of an afterglow.
\section{Monte Carlo simulations: afterglows of long GRBs}
\label{s:simul}

As was described in \S\ref{s:tasks}, the \emph{SRG} all-sky survey
will not be homogeneous: the total accumulated exposure (per source)
will be significantly larger at high ecliptic latitudes relative to the 
ecliptic plane. The ecliptic caps will also be advantageous for
studying X-ray transients, because a point source will pass
through the eROSITA FoV during more than 6 consecutive rotation cycles
and can thus be monitored for significantly longer than 1~day. We 
performed Monte Carlo simulations of the 'GRB X-ray afterglow sky' as 
would be seen by eROSITA, taking the planned survey strategy into
account. The simulation consisted of the following steps: 1) model the 
position of the eROSITA FoV as a function of time (by calculating the
phases of rotation around the Sun and the satellite's axis)
over the duration of the survey (several years), 2) assume a total
rate, flux distribution and light curve characteristics of afterglows, 
3) use these parameters to draw afterglows in random celestial
positions, 4) check if a given event falls into the eROSITA FoV
and satisfies the detection criteria for Task~1 or Task~2 specified
above. 

In what follows, we concentrate on classical
long ($>1$~s) GRBs. Possible additional contributions of other 
classes of bursts are considered in \S\ref{s:other}. 

\subsection{Population of afterglows}

The most important and non-trivial part of the simulation is the
definition of the intrinsic characteristics of the afterglow
population. The problem here is that all samples of afterglows
reported in the literature are probably biased relative to the true
population of such events, because they are based on samples of GRBs 
triggered above certain detection thresholds. In fact, the eROSITA
all-sky survey might help reveal the intrinsic properties of the GRB
afterglow population. 

\subsubsection{Distribution of GRB prompt emission fluences}

The largest existing GRB sample has been provided by the Burst and
Transient Source Experiment (BATSE) on board the \emph{Compton Gamma
  Ray Observatory} (\emph{CGRO}) \citep{stern01}. The $\log N$--$\log
P$ (here $P$ is the peak photon flux in the 50--300~keV energy band) 
distribution of long bursts from this sample, corrected for
the survey's efficiency function, can be described by a number of
models (\citealt{stern02}, hereafter S02). In our simulations, we used 
simple approximations to some of these models (namely to SF1,M and
SF3,M from Table~2 in S02) and their extrapolations to lower
fluences. Specifically, our 'minimal' model M1 reproduces the measured 
$\log N$--$\log P$ distribution of bursts above the BATSE detection
limit and has a sharp cut-off below this threshold; this model has
nearly the same normalisation (total GRB rate) as the SF1,M model in 
S02. Our 'medium' model M2 is the same as M1 but extrapolated by a
constant level to lower fluences (down to $10^{-8}$~erg~cm$^{-2}$); it
has nearly the same normalisation as the SF3,M model in S02 and 
corresponds to increasing star formation at large ($z>2 $) redshifts
(see \citealt{PorcianiMadau2001}). Finally,  
our 'maximal' model M3 is the same as M2 but has an additional
power-law component with a slope of 3/2 (also extending down to 
$10^{-8}$~erg~cm$^{-2}$), whose normalisation is chosen so as to
distort the observed BATSE distribution near its threshold by
$\sim$~20\%, just within the uncertainty of this distribution (see
e.g. \citealt{Sazonov2004}). Such an additional component may be
present due to the existence of low-luminosity GRBs
(e.g. \citealt{Kulkarni1998, Soderberg2004}). Although only a few such
events have been detected and associated with nearby ($z\lesssim 0.1$)
supernovae so far, the total rate of low-luminosity GRBs in the
Universe can be much higher than that of classical (high-luminosity)
GRBs \citep{Pian2006,Liang2007} and such events occurring at moderate
distances ($z\lesssim 1$) can emerge in  large numbers at currently
inaccessible low fluences. 

\begin{figure}
\centering
\epsfxsize=1.0\columnwidth \epsfbox{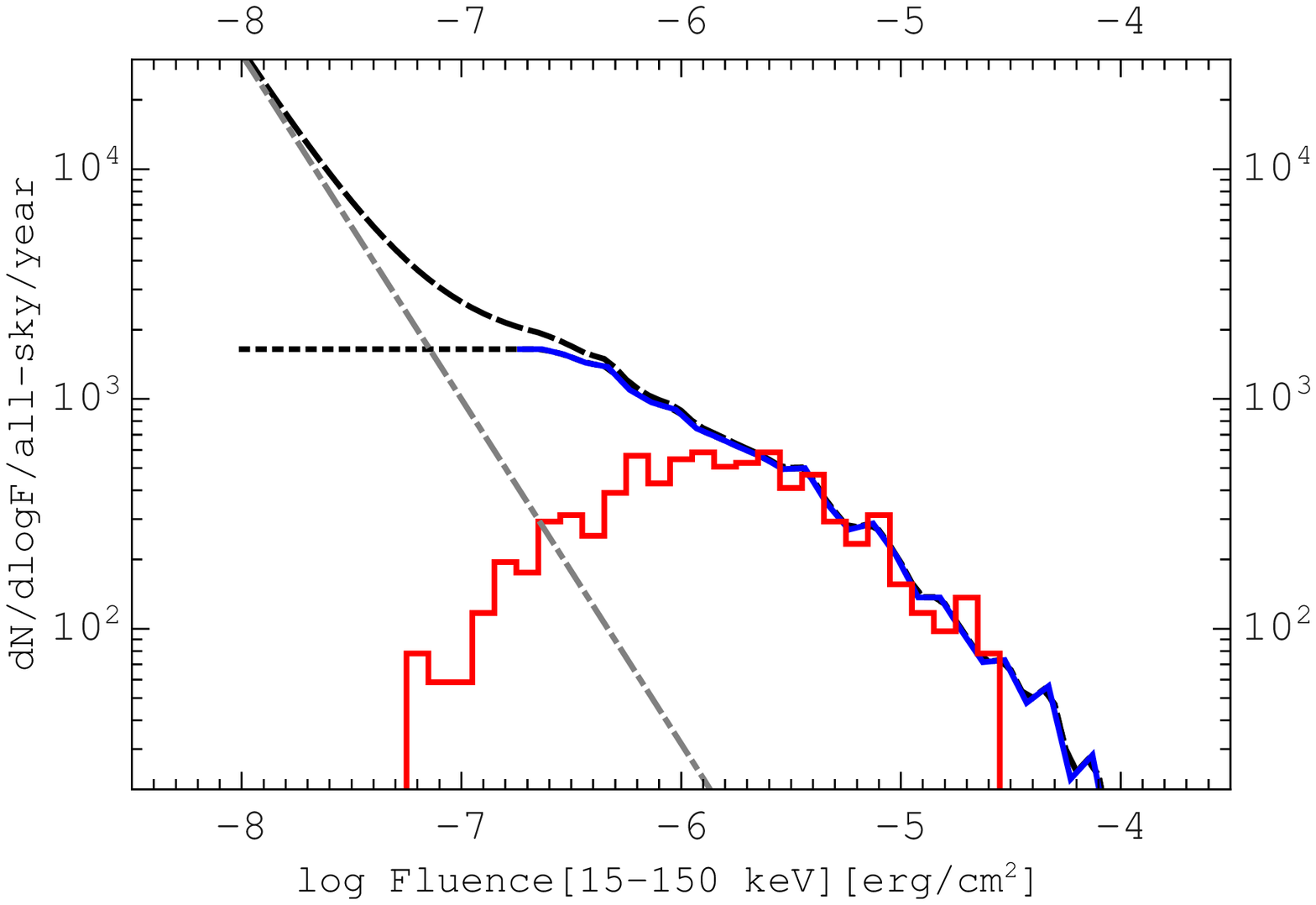}
\topcaption{GRB number--fluence (15--150~keV) distribution. Blue --
  the measured distribution of \emph{CGRO}/BATSE bursts, corrected for
  the detection efficiency function \citep{stern01} and converted from
  peak fluxes to fluences (see text). Black -- various plausible 
  extrapolations of the BATSE distribution to lower fluences: dotted
  line -- flat extrapolation, dashed line -- flat extrapolation plus
  an additional power-law component with a slope of 3/2 (also shown 
  separately by the gray dash-dotted line). Also shown is the observed 
  (and thus distorted at low fluences by the detection efficiency
  function) distribution of \emph{Swift}/BAT bursts (red). }
\label{logNlogSf}
\end{figure}

To proceed to the statistical properties of GRB afterglows, we need to 
convert the BATSE distribution ($\log N$--$\log P$) of GRB peak fluxes
in the 50--300~keV energy band to a distribution of GRB fluences in
the \emph{Swift}/BAT energy band (15--150~keV). To this end, we first
convert peak photon fluxes from 50--300~keV to 15--150~keV by the
method described in \cite{dai09}, which provides rather good agreement
between the BATSE and BAT $\log N$--$\log P$ distributions, and then 
the 15--150~keV peak photon fluxes to 15--150~keV fluences S$_{15-150}$
using the average ratio of these two quantities found for the BAT
sample \citep{Sakamoto11}. Fig.~\ref{logNlogSf} shows the resulting
distributions (hereafter $\log N$--$\log S$) for the different GRB
population models described above.  

\subsubsection{Distribution of GRB afterglow fluxes}

Comprehensive studies of GRB X-ray afterglows have been done with
BeppoSAX \citep{dePasquale06} and \emph{Swift}/XRT
\citep{Sakamoto08,gehrels08}. Due to the higher sensitivity of the
latter, \cite{berger05} argued that the \emph{Swift}/XRT sample has
the least bias, hence we base our analysis on the properties of this
sample.  

\begin{figure}
\centering
\epsfxsize=1.0\columnwidth \epsfbox{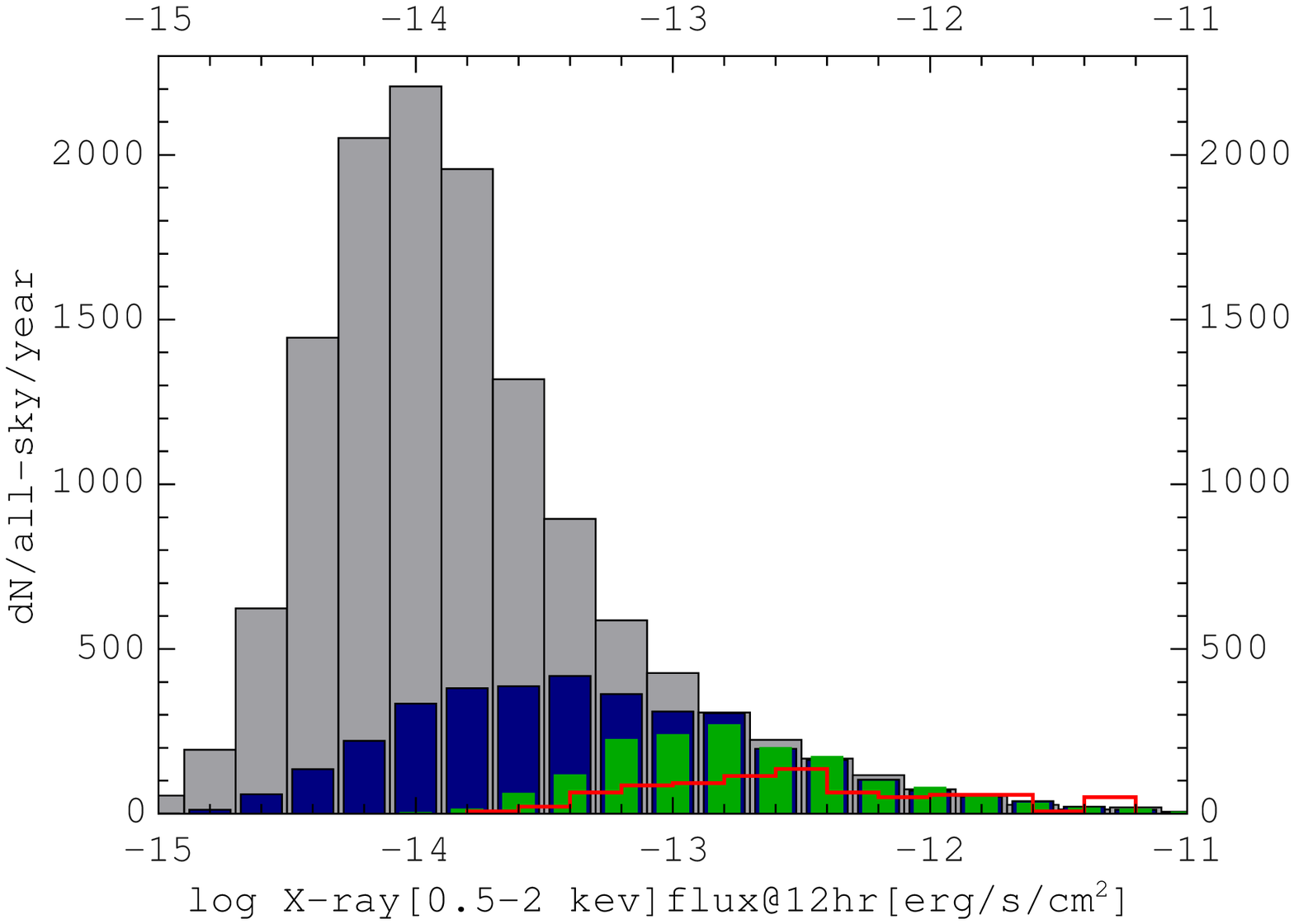}
\topcaption{Distribution of X-ray afterglow fluxes: red -- measured by
  \emph{Swift}/BAT, green -- simulated for model M1 (cut-off at
  low fluences), blue -- simulated for model M2 (flat extrapolation to
  low fluences), gray -- simulated for model M3 (flat extrapolation
  and additional power-law component with a slope of 3/2 at low
  fluences). The decrease at the low-flux end in the simulated
  distributions is due to the truncation of the original GRB samples
  at $10^{-8}$~erg~cm$^{-2}$.}
\label{xfluxdistr}
\end{figure}

We need to make a conversion from the distribution of GRB prompt
emission fluences ($S$) discussed above to a distribution of X-ray
afterglow fluxes (e.g. at 12~hours after the prompt emission,
$F_{X,12}$). A strong linear correlation was found between $S$ and
$F_{X,12}$ for \emph{Swift}/XRT GRBs \citep{Sakamoto08,gehrels08}. We
thus use this correlation to predict $F_{X,12}$ for a given $S$ but
also take into account the scatter associated with the
correlation. Specifically, we distribute $F_{X,12}$ lognormally with 
variance $\sigma=0.31$ around the expected value. In addition, we make 
a correction to the 0.5--2~keV energy band from the \emph{Swift}/XRT
0.3--10 keV  band) assuming a power-law spectrum with $\Gamma=2$ and
low X-ray absorbing column density $\NH\leq 10^{22}$~cm$^{-2}$, as is
typical of GRB afterglows \citep{campana11}. Fig.~\ref{xfluxdistr}
shows the resulting distributions of afterglow fluxes for the M1, M2
and M3 models described above.

\subsubsection{Afterglow light curves}

To finally simulate the `eROSITA afterglow sky', we need to specify
the shape of afterglow light curves. Measured afterglow decay indices
are distributed in a narrow range around $\delta=1.3$
\citep{berger05}, hence we simply fix the light curve slope at this
value. 

\subsection{Results}

\begin{figure}
\centering
\includegraphics[width=1.0\columnwidth]{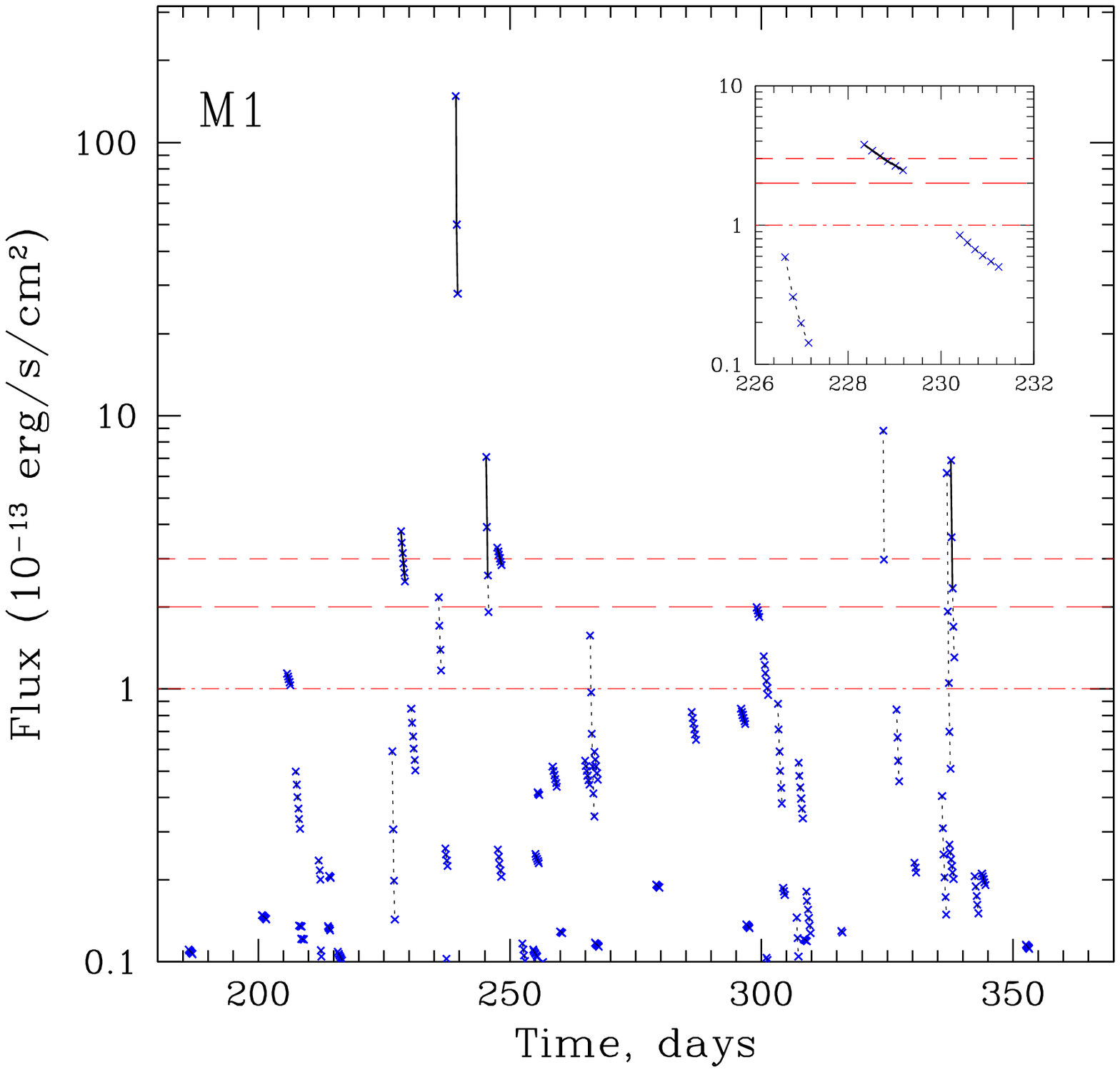}
\topcaption{Simulated eROSITA half-a-year record history of GRB
  afterglows for the M1 (minimal) model of GRB population. The red 
  lines indicate different eROSITA detection limits (see text) per
  scan. Successive detections of the same afterglow are connected 
  with dotted lines. Those light curves that satisfy the
  identification criteria (at least three flux measurements higher
  than $F_e = 2 \times 10^{-13}$~erg~s$^{-1}$~cm$^{-2}$) are
  shown by solid lines. The inset shows a few afterglows with better
  time resolution. 
}
\label{lcurve}
\end{figure}

Fig.~\ref{lcurve} demonstrates an example of a simulated eROSITA
half-a-year all-sky scan for model M1. Table~\ref{sum}
summarises the detection rates for Task~1 and Task~2 for the different
considered models, obtained by averaging over a large number
of simulated all-sky scans.

The simulated rates are somewhat higher than the analytical lower
limits presented in \S\ref{s:tasks} due to i) the contribution 
of weak ($F_{X,12}<F_e$) events neglected in the analytical
calculation (only relevant for the $N_1$ rate, see
\S\ref{s:tasks}) and ii) the increased number of scans at high
ecliptic latitudes. The differences in the rates obtained for
various models (M1, M2 and M3) are associated with the contribution of  
low-fluence GRBs. The $N_2$ rate is especially sensitive to such
events, because eROSITA with its effective detection threshold ($F_d$)
may find a large number of afterglows from weak GRBs that are below
the detection threshold of the current and previously flown GRB
detectors. 


\begin{table}
\topcaption{Summary of the results of simulations for afterglows of
  long GRBs. Only GRBs with fluence higher than
  $10^{-8}$~erg~cm$^{-2}$ (15--150~keV) were taken into account. For
  $N_1$, column \emph{a} is for the detection limit $F_{e}=3\times
  10^{-13}$~erg~s$^{-1}$~cm$^{-2}$, while column \emph{b} is for
  $F_{e}=2\times 10^{-13}$~erg~s$^{-1}$~cm$^{-2}$. The detection
  limit for $N_2$ is $F_{d}= 1 \times 10^{-13}$~erg~s$^{-1}$~cm$^{-2}$.} 
\centering
\begin{tabular}{ccccc}\hline\hline
\multirow{2}*{Model}&{Normalisation,}&\multicolumn{2}{c}{
  $N_1$, per year}&\multirow{2}*{$N_2$, per year} \\ 
{} &{all-sky per year}&{a}&{b}&{}\\
\hline
{M1}&{1600}&{4.4}&{6.8}&{19.5}\\
{M2}&{3600}&{4.8}&{7.2}&{27.4}\\
{M3}&{3600+9100}&{5.0}&{7.6}&{55.6}\\
\hline\\
\end{tabular}
\label{sum}
\end{table}

\section{Confusion with other types of X-ray sources}
\label{s:cont}

Our proposed algorithm for detection of afterglows of non-triggered GRBs 
(Task~1) is based on checking if three successive X-ray flux
measurements separated by 4~hours are consistent with a power-law
decline. However, other types of astrophysical objects detected 
during the \emph{SRG}/eROSITA all-sky survey might also exhibit
similar variability properties and could thus be misidentified as
GRB afterglows. Below we discuss the most important potential
contaminants and how they can be distinguished from afterglows.
 
\subsection{Active Galactic Nuclei}

A few millions of  active galactic nuclei (AGN) are expected to be
discovered by eROSITA all over the sky during the 4-year survey
(e.g. \citealt{Predehl11}). Although less than $\sim 10^5$ of these
will be brighter than $F_e=2\times 10^{-13}$~erg~s$^{-1}$~cm$^{-2}$
(0.5--2~keV) -- our effective detection limit for GRB afterglows, it
is still an enormous number of sources, some of which may be confused
with GRB afterglows. Indeed, AGN are known to be significantly
variable on various time-scales from minutes to years (see
\citealt{McHardy10} for a review) and usually have Crab-like X-ray
spectra with $\Gamma\sim 2$, similar to the spectra of GRB afterglows.  

The problem of AGN X-ray variability is far from being completely
explored and will be one of the key scientific topics of the
\emph{SRG} mission. Nevertheless, the results of previous
studies allow us to roughly estimate the number of AGN 
that might resemble GRB afterglows in the eROSITA survey. 
\emph{Rossi X-ray Timing Explorer} (\emph{RXTE}) observations 
have demonstrated that variability properties of AGN are similar to
those of Galactic black-hole X-ray binaries (BHB), except that the
characteristic time-scales are longer in proportion to the black hole   
mass \citep{McHardy04}. Typically, X-ray light curves of AGN are
characterised by a power spectrum density (PSD) that can be fitted by 
a power law with a slope of $-2$ at high frequencies and a
slope of $-1$ below some frequency $\nu_B$. This characteristic
frequency correlates with the black hole mass $M_{BH}$ and accretion
rate $\dot{m}_E$ (in units of the Eddington critical rate)
\citep{McHardy10} and thus varies significantly between different
AGN. Assuming a break frequency $\nu_B=10^{-4}$~Hz and rms/flux ratio
of 0.2 (typical values for AGN, see \citealt{McHardy10} and references
therein), we simulated a large number of AGN light curves using the
method described by \cite{Timmer95} and then applied our GRB afterglow
detection criterion (Task~1) to these light curves. We found that the
probability for an AGN light curve to have three successive eROSITA
flux measurements (separated by 4~hours) that are consistent with a
power-law decay with $0.5<\delta<3.5$ is $\sim 1$\%. However, in
approximately 3/4 of such sequences, the subsequent (fourth) flux will
be higher than the third one, which will clearly indicate against a
GRB afterglow. 

Therefore, the absolute majority of episodes of significant
variability in AGN will not resemble GRB afterglows in eROSITA
observations. Nevertheless, we can expect a total of a few hundred
episodes associated with AGN that will mimic GRB
afterglows. Fortunately, most of such bright, $F_X>F_e=(2-3)\times
10^{-13}$~erg~s$^{-1}$~cm$^{-2}$ (0.5--2~keV), AGN will manifest 
themselves as persistent sources (and hence exclude an afterglow
origin) over the course of the mission, as they will be repeatedly
detected in successive (every 6 months) eROSITA all-sky scans (the
detection limit for a point source in one all-sky scan is $\sim
7\times 10^{-14}$~erg~s$^{-1}$~cm$^{-2}$, 0.5--2~keV, see Appendix 2). Most of such
AGN will also be identified as such through multi-wavelength follow-up
efforts (see \S\ref{s:optical} below). 

\subsection{Stellar flares}

Coronally active stars are another population of variable sources that
could produce afterglow-like light curves. Indeed, several of the 23
afterglow candidates in the RASS sample of \cite{Greiner00} showed
power-law-like light curves (constructed similarly to the
anticipated eROSITA data, namely there were several 10--30~s exposures
separated by 1.5-hour intervals) but proved to be associated with
stars.  

Stellar flares demonstrate a remarkable variety of properties such as
the duration and shape of the rising and decaying phases,  
the peak flux and its ratio to the quiescent level. Unfortunately, 
virtually all studies of these properties reported in the 
literature are strongly affected by the detection efficiency and
strategy of the particular surveys (see \citealt{Favata2003,Guedel2004} 
for review). Therefore, it is difficult to make even crude estimates of 
the frequency of stellar flares resembling GRB afterglows in eROSITA 
observations. Nevertheless, we may mention a couple of aspects
that may help distinguish stellar episodes from afterglows. First,
there is a well-known tendency for stonger flares to occur in stellar
coronae with stronger quiescent X-ray activity
\citep{Favata2003,Guedel2004}). Therefore, a flare exceeding our
proposed threshold for GRB afterglows, $\sim 2\times
10^{-13}$~erg~s$^{-1}$~cm$^{-2}$ (0.5--2~keV), should typically be
associated with a star that will be seen as a persistent source (with
$F_X\gtrsim 5\times 10^{-14}$~erg~s$^{-1}$~cm$^{-2}$, 0.5--2~keV) in
repeated eROSITA all-sky scans. Secondly, typical stellar flares have
spectra that are typically significantly softer \citep{Favata2003}  than those of
GRB afterglows. Although our afterglow detection limit is relatively
low ($\sim 9$~counts during the third flux measurement), the total
number of photons accumulated during the $f_1$ measurement will
typically be $\sim 10^2$ in the 0.5--2~keV energy band, so that there
will be a sufficient number of photons detected above 2~keV to
discriminate the hard spectrum of an afterglow from the softer
spectrum of a stellar flare. Additional valuable information 
above 6~keV can be provided by the ART-XC telescope if an afterglow
falls into its (smaller relative to eROSITA) FoV. Finally, as we
discuss below, multi-wavelength information should be very helpful. 
 
\subsection{Cross-correlation with optical and infrared catalogues}
\label{s:optical}

The anticipated availability by the time of the \emph{SRG}
mission of a number of sensitive optical and infrared photometric
surveys covering the whole sky or its large fractions should greatly
help in distinguishing coronally active stars and AGN from host
galaxies of GRB afterglows. Indeed, the non-flaring X-ray to optical
flux ratio, $F_X/F_{\rm opt}$, never exceeds $\sim 10^{-3}$, for
stellar coronae (e.g. \citealt{Guedel2004}), and only during the most
extreme flares can the coronal X-ray luminosity become comparable to
the bolometric luminosity of the star (see \citealt{Osten2010} and
references therein; also \citealt{Uzawa2011}). Therefore, if a 
stellar flare satisfies our GRB afterglow detection criterion (Task~1)
and hence have a peak flux $F_X\sim 10^{-12}$~erg~s$^{-1}$~cm$^{-2}$
(0.5--2~keV), it should be readily associated with a relatively bright
star of $R\lesssim 20$ in the Sloan Digital Sky Survey (SDSS, which
has imaged $\approx 1/3$ of the sky in the $u$, $g$, $r$, $i$ and $z$
bands with a typical sensitivity of $R=22.2$, Data Release 8,
\citealt{Aihara2011}) and/or in the 3$\pi$ Steradian Survey of the the
Panoramic Survey Telescope and Rapid Response System (Pan-STARRS, which
is planned to image 3/4 of the sky in the $g$, $r$, $i$, $z$ and $y$ 
filters down to $R=24$, \citealt{Kaiser2002}). Indeed, the number
density of $R<20$ stars at high Galactic latitudes is less than 1~per
square~minute (e.g. \citealt{Juric2008}), whereas the eROSITA
localisation accuracy is better than 10~arcsec. 

As concerns AGN, based on the results of a number of previous 
extragalactic X-ray surveys (e.g. \citealt{Aird2010}) it
can be expected that the optical to X-ray flux ratio will be in the
range $0.1 <F_X/F_{\rm opt}<10 $ for the majority of AGN detected by
eROSITA. Consequently, relatively bright AGN with $F_X\gtrsim 
10^{-13}$~erg~s$^{-1}$~cm$^{-2}$ (0.5--2~keV), i.e. such AGN that may
produce flares resembling GRB afterglows, should be associated with
relatively bright, $R\sim 19\pm 2$, objects in the SDSS and
PanSTARRS catalogues. Furthermore, it is expected (Sazonov et al., in
preparation) that the absolute majority of such moderately X-ray
bright AGN should be readily identifiable with the help of the
recently published all-sky catalogue of mid-infrared sources detected by
the \emph{Wide-Field Infrared Survey Explorer} (\emph{WISE},
\citealt{Lake2012}).   

In comparison, the host galaxies of typical GRBs observed so far are 
generally faint, with about half being weaker than $R_{ab}=23.5$ and
$K_{ab}=22.5$ \citep{Savaglio2009}. Since, there is no significant dependence of GRB fluence on redshift \citep{Sakamoto11}, the host galaxies of GRBs associated with eROSITA afterglows, will typically be as dim. We conclude that the absence of a bright counterpart in the publicly available catalogues of large-area optical and infrared
photometric surveys will be a strong additional signature in favour
of a GRB afterglow origin of transients showing a power-law-like
decline. 

\section{Afterglows from other classes of GRBs}
\label{s:other}

Some uncertainty in the number of afterglows that can be
detected with eROSITA arises from the lack of information about the
low-fluence end of the $\log N$---$\log S$ distribution of long
GRBs. In addition, we should take into account the possibility of
detecting afterglows associated with some additional subclasses of
GRBs. 

\subsection{Short GRBs}

In comparison to long GRBs (LGRB), short GRBs (SGRB) are rarer and
fainter. Indeed, about 10 long bursts are localised by  
\textit{Swift} for every short burst, and the average fluence of SGRBs
is an order of magnitude lower than that of LGRBs. 
The X-ray afterglows of SGRBs are also faint, which might point at
different properties of the progenitor and circumburst 
environment. However, using a large sample of 37 short
and 421 long GRBs, \citet{Nysewander2009} found that SGRBs follow
approximately the same correlation between X-ray afterglow brightness
and prompt $ \gamma $-ray fluence as do LGRBs. This suggests that the
eROSITA detection rate of SGRB afterglows should be relatively
low. Adopting $F_{e}=2 \times 10^{-13}$~erg~s$^{-1}$~cm$^{-2}$ as the
detection limit for Task~1, we find that only 7 of 27 SGRBs with X-ray
afterglow flux measurements in the sample of \cite{Nysewander2009}
have $ \tau_m >0$, i.e. $\tau_e >8$~hr, implying that the
corresponding detection rate $N_1$ of SGRB afterglows will be $\approx
40$ times lower than for LGRB afterglows (since almost all LGRB afterglows in the \textit{Swift} sample have $\tau_e >8$~hr). Thus, taking into account our
estimates for long bursts (Table~\ref{sum}), SGRB afterglows will only 
rarely, $\sim 0.1$~per~year (see Table~\ref{typesum}), satisfy the
eROSITA detection criterion for non-triggered GRBs. 
 
\begin{table}
\centering
\topcaption{Expected detection rates for afterglows
  associated with various types of GRBs}
\begin{tabular}{cc}\hline\hline
{Type}&{N$_1$ ($F_{e}= 2 \times 10^{-13}\frac{{\rm erg}}{{\rm
      s}~{\rm cm}^{2}}$), per year}\\ 
\hline
{Long GRBs}&{ 6.8  --  7.6}\\
{Short GRBs}&{$\sim$ 0.1}\\
{GeV GRBs}&{$ < $ 1.5}\\
{XRFs}&{$\sim$ 0.4}\\
\hline\hline\\
\end{tabular}
\label{typesum}
\end{table}

\subsection{Ultra hard (GeV) GRBs}

The energy range of \emph{Swift}/BAT is narrow (15--150~keV) with
respect to the observed variety of GRB prompt emission  spectra
\citep{Virgili2011}. The effective energy range of \emph{CGRO}/BATSE
for GRB detection was similar, 50--300~keV. This suggest that there
might exist a population of GRBs with ultra-hard prompt 
emission and moderate or undetectable signal in the hard X-ray
band. This hypothesis can now be tested with the Large Area Telescope
(LAT) on board \textit{Fermi}, which is sensitive from 30~MeV to
100~GeV and has a FoV of $ \approx 2.4$~sr \citep{Atwood2009}. 
The detection rate of \textit{Fermi}/LAT is $\approx 10 $ bursts per
year, which corresponds to $R \approx 50 $ bursts per year over the
whole sky. All of the long GRBs detected by LAT so far occurred
outside the BAT FoV. Detailed analysis of prompt and afterglow
emission properties of the LAT-detected GRBs \citep{Racusin2011,
  Cenko2011} indicates that although these GRBs have the largest fluences
ever observed, their afterglows still fall within the flux
distribution of \emph{Swift}/BAT-triggered GRBs, exactly near
its upper boundary, whereas the spectral and temporal behaviour of the 
LAT-triggered GRB afterglows are quite similar to those of the 
BAT-triggered ones. 

Because the scatter in the X-ray afterglow fluxes of LAT-triggered
GRBs is fairly small (probably owing to the fairly poor statistics
provided by less than 10 GRBs in the current sample), some   
instructive estimates can be obtained assuming that all of these
afterglows have some 'average' form. Using the data from
\citet{Swenson2010}, we find that a typical LAT-triggered GRB
afterglow has $ F_{X,12}\sim 2 \times 10^{-12}$~erg~s$^{-1}$~cm$^{-2}$
(0.5--2~keV flux at 12 hours after the prompt emission) and $
\alpha\approx 1.5 $ (temporal decay index). Thus, assuming an eROSITA
detection limit per scan of $F_{e}= 2 \times
10^{-13}$~erg~s$^{-1}$~cm$^{-2}$, such afterglows can be detected by
eROSITA up to $\tau_e\approx 57$~hours after the prompt emission (see
Sect.~\ref{s:task1}). Hence, the corresponding detection rate
$N_1\approx 0.66$ per year over the whole sky (given a total rate of
$R = 50$ of such bursts per year). In addition, there might be faint
ultra-hard GRBs remaining below the LAT detection limit but capable of
producing X-ray afterglows detectable by eROSITA for more than 8
hours. However, by analogy with classical GRBs (see Table~\ref{sum}),
the number of such afterglows is unlikely to exceed the number of
afterglows associated with detectable ultra-hard GRBs. We thus expect
the total eROSITA detection rate of afterglows of ultra-hard GRBs to
be less than 1.5~per~year (see Table~\ref{typesum}). 

\subsection{X-ray Flashes}

There can also be X-ray afterglows associated with so-called 'X-ray
flashes', originally burst-like (shorter than 1000~s) events detected
in X-rays by the Wide Field Camera on \emph{BeppoSAX} but not detected
by the Gamma-Ray Monitor on the same satellite
\citep{Heise2001}. Comprehensive studies based on data from
\emph{BeppoSAX} \citep{Kippen2003}, \emph{HETE-2} \citep{Sakamoto2005}
and most recently \emph{Swift} \citep{Sakamoto2008a} have not found
significant differences between the duration and 
sky distributions of XRFs and 'classical' GRBs. Moreover,
the spectral properties of XRF prompt emission were found to be
similar to those of GRBs, except that the peak energies
$E_{peak}^{obs}$ (of the prompt $\nu F_\nu$ spectrum), peak fluxes
$F_{peak}  $ and fluences $ S_E $ of XRFs are much smaller. This
suggests that X-ray flashes arise from the same phenomenon
as GRBs, continuing the GRB population to low $E_{peak}$
\citep{Sakamoto2008a}.   

The Wide-Field X-ray Monitor on board \emph{HETE-2} makes it possible
to readily estimate the rate of observable XRFs in comparison with
GRBs, because its threshold for detection and localisation of bursts
in terms of the peak photon number flux is only weakly dependent on
$E_{peak}^{obs} $, i.e. nearly the same for XRFs and GRBs 
\citep{Sakamoto2005}. Based on the fluence ratio $S_X(2-30~{\rm
  keV})/S_\gamma(30-400~{\rm keV})$, out of a total of 45 bursts in the
\emph{HETE-2} sample there are 16~XRF, nearly half of which have
$E_{peak}^{obs}<20$~keV and a peak flux of $<0.2$~ph~s$^{-1}$~cm${^2}$
in the 50--300~keV energy band. Here we are interested in such events
because i) they seem to be missing from the $\log N$--$\log S$
distribution of BATSE GRBs (see figures 16 and 17 in
\citealt{Sakamoto2005}), and ii) detection of such bursts by
\emph{Swift}/BAT in its 15--150~keV energy range is
challenging. Studies of XRF afterglow emission based on
\emph{BeppoSAX} and \emph{HETE-2} \citep{deAlessio2006} and on
\emph{Swift} \citep{Sakamoto2008a} samples have produced somewhat
controversial results. In particular, \cite{deAlessio2006} found that
XRF afterglow light curves are similar to those of classical GRBs,
including the break feature at late times \citep{Panaitescu2007}. The
XRF afterglow fluxes were also found to be not much lower than for GRB
afterglows. On the other hand, the afterglows of XRFs from the more
extensive \emph{Swift}/XRT sample, consisting of bursts with
$E_{peak}^{obs} > 20$~keV, exhibit no break in the light curves and  
are weaker by a factor of 2 or more compared to that of classical GRBs
\citep{Sakamoto2008a}. To extend this result to XRFs with
$E_{peak}^{obs} < 20$~keV, we may assume that there is a positive
correlation between prompt emission fluence and X-ray afterglow flux,
i.e. the afterglows of such XRF should be fainter than those of XRFs
with $E_{peak}^{obs}>20$~keV. Putting all these facts together, we
estimate that afterglows of XRFs with $E_{peak}^{obs}<20$~keV are
approximately 4 times rarer and at least 2 times fainter than
afterglows of typical \emph{Swift} bursts. Therefore, the total 
detection rate of the former is expected at the level $\sim 0.4$ per year (see Table~\ref{typesum}).   
\section{`Orphan' afterglows and failed GRBs}

Even after considering various subclasses of GRBs and XRFs, our
account of afterglows that can be detected by eROSITA is still
incomplete. An additional contribution may come from afterglows that
are not preceded by prompt $\gamma$- or X-ray emission, which are
often referred to as 'orphan afterglows' (OA). The standard theory of
GRBs, dealing with strongly beamed emission of a highly
relativistic (initial Lorentz factor $\Gamma \sim 100-1000$) jet,
allows for the existence of three classes of OA: 1) on-axis 
afterglows, observed within the opening angle of the initial
relativistic jet but outside its narrow $\gamma$-ray emitting (due
to a larger Lorentz factor) component; 2) off-axis  
afterglows, observed outside the initial jet after the jet break time,
when the jet expands sideways \citep{Nakar2003}; and 3) afterglows of
so-called 'failed GRBs' (FGRB), associated with baryon-contaminated
fireballs with an initial Lorentz factor of much less than 100 but
still greater than unity \citep{Huang2002}. 

The light curve of an off-axis OA should initially rise until a moment
that approximately corresponds to the jet-break time of 'normal' GRB
afterglows ($10^4-10^5$~s, \citealt{Panaitescu2007}) and then decay in
the usual post-jet-break manner (with a decay slope greater than 1.5
and typically near 2.0, see \citealt{Nakar2003}). Hence,
for such an afterglow to satisfy our Task~1 identification criteria
(i.e. a power-law-like decay), the required three
successive (spanning 8~hours) detections by eROSITA should take place
after the moment of maximal flux. However, the X-ray flux at
this stage will typically be already too low for detection. Therefore, such
off-axis events are unlikely to be found in significant numbers by
eROSITA. As regards on-axis OAs, \citep{Nakar2003} summarised 
observational constraints provided by the numbers of X-ray transients
and GRBs detected by \emph{Ariel~5}, \emph{HEAO-1}, \emph{ROSAT} and
\emph{BeppoSAX} and concluded that the average ratio of the opening
angle of GRB jets to that of their $\gamma$-ray emitting components is
$\lesssim 2$. This suggests that eROSITA may find a comparable 
number of orphan afterglows to that of 'normal' afterglows,
i.e. several events per year (see Table~\ref{sum}) and can thereby
improve the constraints on the structure of relativistic jets in GRBs.   
 
As concerns failed GRBs, their existence seems rather natural from a
theoretical point of view, because it is hard to produce a pure (in
terms of the baryonic content) highly relativistic flow. The
collapsar model \citep{MW1999} predicts a great variety in the baryon
mass and released energy, and consequently in the initial Lorentz
factor $\Gamma$ of the produced fireball, among different events. In
most cases, $\Gamma$ is expected to be low ($ \ll 100 $), leading to
FGRBs that have almost the same initial energy as normal GRB
fireballs ($10^{51}-10^{53}$~erg) but are polluted by baryons with
mass $\sim 10^{-5}-10^{-3} M_{\odot}$. It has also been suggested that
a large population of FGRBs could arise if the jet 
failed to break out of the progenitor star in the collapsar model
\citep{Bromberg2011}. In such a case, a UV/soft X-ray thermal burst
followed by an afterglow with some distinct features should be
observed \citep{Xu2012}. It is practically impossible to predict the
rate of such events, but eROSITA may shed light on this explored
population.   

\section{Summary and discussion}

The results of this study imply that it should be possible to detect
and identify, by the shape of the light curve, 4--8 X-ray afterglows
associated with classical long GRBs per year in the eROSITA all-sky
survey data (see Table~\ref{sum}). The exact number will depend
on the shape of the $\log N$--$\log S$ distribution of GRBs at low
fluences (near the effective threshold of \emph{CGRO}/BATSE and
\emph{Swift}/BAT.  In addition, eROSITA is expected to find a small
number of afterglows associated with other classes of GRBs such short
bursts, GeV bursts and X-ray flashes (Table~\ref{typesum}). Thus, by
the end of the 4-year survey, a sample of at least 20--40 X-ray
afterglows can be accumulated. This sample, although smaller than the
already existing samples of afterglows, will nevertheless be
interesting for systematic studies of GRBs and their afterglows
because of its unbiased nature. In particular, it can be used to
construct an unbiased distribution of X-ray afterglow fluxes and to
obtain constraints on the shape of the $\log N$--$\log S$ distribution 
of GRBs. 

The total number of afterglows detected by eROSITA may prove higher,
perhaps by a factor of 2 or more, due to orphan afterglows and failed
GRBs. The actual detection rate will thus provide interesting
constraints on the properties of relativistic jets associated with the
collapse of massive stars.

The proposed algorithm for searching afterglows of non-triggered GRBs
(Task~1) in the eROSITA data is based on checking if the light curve of a
given source resembles a power-law decline. Although such a
procedure may erroneously identify a large number of other types of
variable X-ray sources (e.g. AGN and stellar flares) as GRB
afterglows, most of such contaminants should be easily revealed
through cross-checking of successive eROSITA all-sky scans and
cross-correlation with large-area optical and infrared source
catalogues.   

We have also discussed the possibility of using the eROSITA data for
searching for afterglows of triggered (by any GRB monitors)
GRBs. Since the coordinates of the triggered bursts will be known,
eROSITA just needs to detect a few photons from the afterglow in one
$\sim 40$~s scan. As a result, the total number of such events can be
large, $\sim 20-60$~per year depending on the $\log N$-$\log S$
function (see Table~\ref{sum}), provided that at the time of the
\emph{SRG} mission there are GRB monitors covering most of the sky at
any given time. X-ray afterglows detected in this way can be
interesting for statistical studies addressing the same scientific
problems as discussed above in relation to the search for afterglows
of non-triggered GRBs. 

Finally, although \emph{SRG} data transfer is planned to occur only once
per day, accurate ($\lesssim 30''$) coordinates provided by eROSITA within
$\sim 1$~day after the event on afterglows of GRBs and orphan
afterglows can be valuable for scheduling further follow-up  observations. 

\section*{Appendix 1: Task 1 detection limit}

Here we determine the eROSITA 0.5--2~keV detection limit $F_e$ for
candidate GRB afterglows. The question we need to
answer is to what accuracy do we need to obtain three consecutive
flux measurements, $f_1$, $f_2$ and $f_3$, to be able to conclude with
certainty that the X-ray flux is decaying as a power law? Because
the result is mostly sensitive to $f_3$ (i.e. the latest flux
measurement), the condition $f_3>F_e$ must be fulfilled. 

\begin{figure}
\epsfxsize=1.0\columnwidth \epsfbox{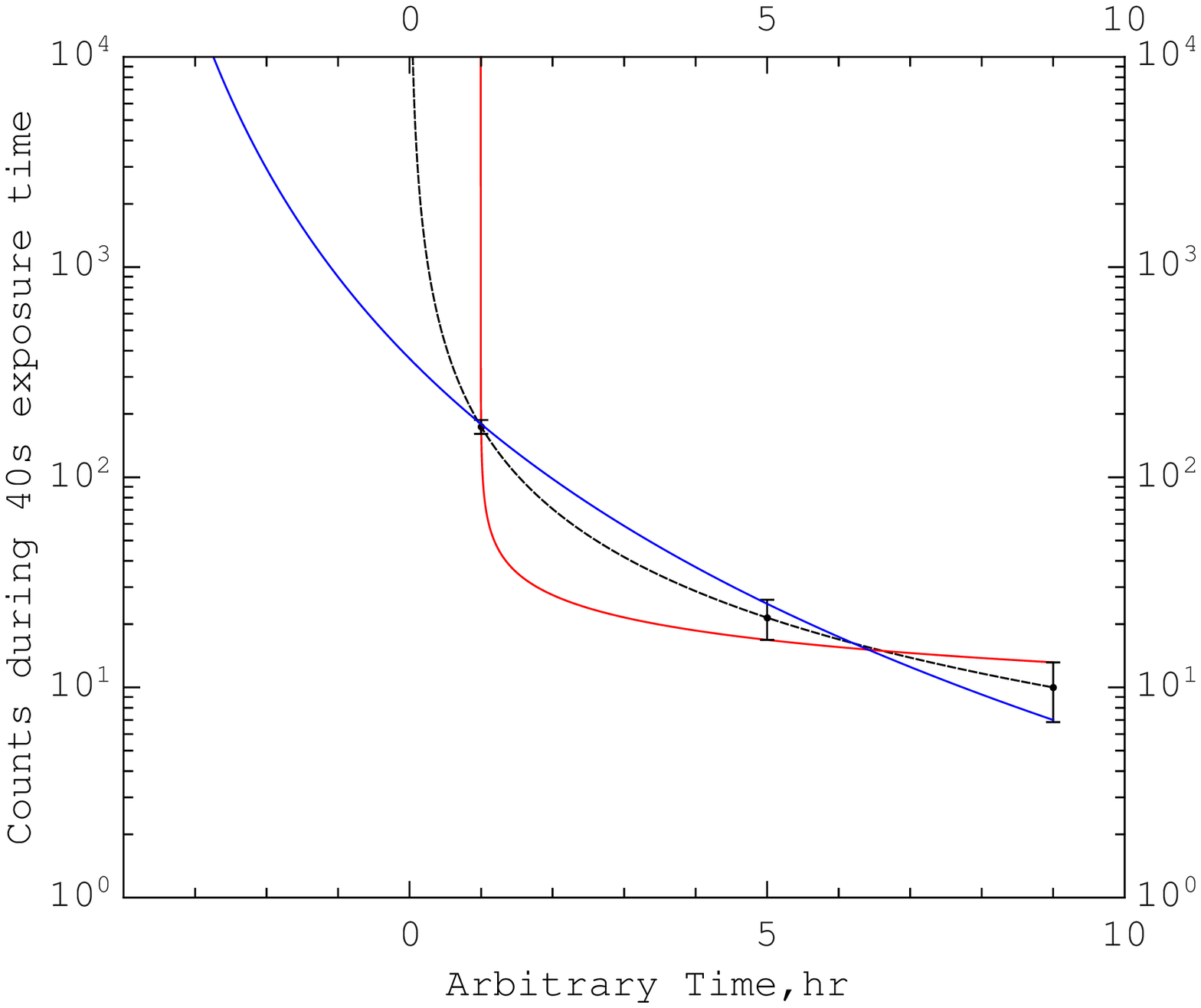}
\caption{Uncertainty in the inferred power-law slope $\delta$ of an
  afterglow light curve induced by the statistical (Poisson)
  uncertainties of count rates measured in 3 consecutive scans. The
  adopted intrinsic light curve is shown in black: its slope is
  $\delta=1.3$, the first detection takes place at $\tau_1=1$~hour
  after prompt emission ($t=0$ here) and the mean flux during the
  third measurement corresponds to 9 photons. For each of a large
  number of simulated data sets, the best-fitting power law was
  found. These solutions are distributed (within $1\sigma$) around the
  intrinsic light curve from $\delta\simeq 0.4$ (red curve) to
  $\delta\simeq 3.3$ (blue curve).}
\label{uncert}
\end{figure}

To this end, we simulated light curves of GRB afterglows as would be
measured by eROSITA, assuming that photon counts obey the Poisson
distribution. The mathematical problem consists of fitting the
intrinsic afterglow parameters $F_{X,12}$, $t_0$ and $\delta$
(equation~\ref{decay}) given $f_1$, $f_2$ and $f_3$ with their 
corresponding statistical uncertainties. In particular, the power-law
slope can be derived analytically from equation~(\ref{delta}). We
found that, if the intrinsic slope $\delta=1.3$ and we 
require the measured value of $\delta$ to lie in the range from
$\approx 0.5$ to $\approx 3.5$ this conservatively broad range
reflects the observed scatter in decay indecies taking into account
the post-break phase \citep{Nysewander2009}), then $f_3$ should be 
measured with accuracy better than $\sim 30$\%, i.e. the latest 
measurement must be based on at least 9 photons. Figure~\ref{uncert}
illustrates this conclusion. It shows the result of multiple
realisations of a light curve for fixed intrinsic parameters:
$\delta=1.3$, $\tau_1=1$~hr (the time of the first flux measurement) 
and $f_3$ corresponding to 9 photons. The resulting family of
possible power-law solutions is bounded (within $1\sigma$) by two
curves that correspond to $\delta \simeq  0.4$ and $\delta\simeq
3.3$. For a characteristic X-ray afterglow spectrum with a photon
index $\Gamma=2$ and the preliminary eROSITA response matrix
(http://mpe.mpg.de/erosita/response/), the 9 counts limit corresponds
  to $F_e\simeq 2 \times 10^{-13}$~erg~s$^{-1}$~cm$^{-2}$ for a
  diametral transition of a source through the eROSITA FoV. 

In reality, the effective detection threshold might be somewhat higher 
than estimated above because there is some uncertainty in the
afterglow spectral properties and characteristics of the eROSITA
telescope. We estimate that $F_e \simeq 3 \times
10^{-13}$~erg~s$^{-1}$~cm$^{-2}$ can be regarded as a conservative value.

\section*{Appendix 2: Task 2 detection limit}

In the case of Task~2, the effective afterglow flux limit is mainly
determined by the background level of the eROSITA detectors.   
It is expected that in the 0.5--2~keV energy band, the cosmic X-ray
background (Galactic and extragalactic)
will dominate over the particle background. After excluding bright
extragalactic sources that will be detected as individual point
sources, the total background count rate in 
the 0.5--2~keV energy band over the eROSITA FoV is expected to $\approx
5$ counts per second \citep{Prokopenko2009}. This
corresponds to $\approx 0.012$ counts inside the region of half-power diameter
(HPD, $\approx 29''$) of the eROSITA point spread function over a
40~s exposure time. According to the Poisson distribution, the
probability that 2 or more background photons will 
be detected by chance in the HPD region is $\approx 7.4\times
10^{-5}$. Assuming that the position of a given GRB is known to an
accuracy of $\sim 10'$ (as is e.g. the case for GRBs detected by
\emph{Swift}/BAT), the probability of finding 2 or more background
events inside this localisation region is less than $3 \times
10^{-2}$, which, however, will not provide 3$ \sigma $ confidence of
source detection. Therefore, it is reasonable to demand at least 3
counts as the detection limit for Task~2. This corresponds to a flux
limit of $F_d\approx 7\times 10^{-14}$~erg~s$^{-1}$~cm$^{-2}$ 
(0.5--2~keV). However, such a detection will provide poor information
about the afterglow brightness. To allow the accuracy of flux
measurement of at least $\sim 50$\% and account for some remaining
uncertainty in the eROSITA background level, we adopt a more
conservative value $F_d= 1 \times 10^{-13}$~erg~s$^{-1}$~cm$^{-2}$,
which approximately corresponding to 4 counts from a point source in
the localisation region.

\section*{Acknowledgements}
The research made use of grants RFBR 11-02-12271-ofi-m and NSh-5603.2012.2, and programme P-21 of the Russian Academy of Sciences. IK acknowledges the support of the Dynasty Foundation.

\end{document}